\documentclass[useAMS,usenatbib,usegraphicx]{mn2e}

\newcommand{\chan}{{\it Chandra}}

\title[X-ray bright sources in the SMC Wing Survey]{X-ray bright sources in the {\it Chandra} Small Magellanic Cloud Wing Survey - detection of two new 
pulsars}
\author[K. E. McGowan et al.]{K. E. McGowan$^{1}$\thanks{E-mail:
kem@astro.soton.ac.uk}, M. J. Coe$^{1}$, M. Schurch$^{1}$, V. A. McBride$^{1}$,
J. L. Galache$^{2}$, 
\newauthor
W. R. T. Edge$^{1}$, R. H. D. Corbet$^{3}$, S. Laycock$^{2}$, 
A. Udalski$^{4}$, D. A. H. Buckley$^{5,6}$ \\
$^{1}$School of Physics and Astronomy, Southampton University, Highfield, 
Southampton, SO17 1BJ \\
$^{2}$Harvard-Smithsonian Center for Astrophysics, Cambridge, MA 02138, USA \\
$^{3}$Universities Space Research Association, X-ray Astrophysics Laboratory, 
Mail Code 662, NASA Goddard Space Flight Center, \\
Greenbelt, MD 20771, USA \\
$^{4}$Warsaw University Observatory, Aleje Ujazdowskie 4, 00-478 Warsaw, 
Poland \\
$^{5}$South African Astronomical Observatory, Observatory, 7935, Cape Town, 
South Africa \\
$^{6}$Southern African Large Telescope Foundation, Observatory, 7935, 
Cape Town, South Africa}

\begin{document}

\date{}

\pagerange{\pageref{firstpage}--\pageref{lastpage}} \pubyear{2006}

\maketitle

\label{firstpage}

\begin{abstract}
We investigate the X-ray and optical properties of a sample of X-ray bright 
sources from the Small Magellanic Cloud (SMC) Wing Survey.  We have detected
two new pulsars with pulse periods of 65.8 s (CXOU J010712.6-723533) and 700 s 
(CXOU J010206.6-714115), and present observations of two previously known 
pulsars RX J0057.3-7325 (SXP101) and SAX J0103.2-7209 (SXP348).  Our analysis 
has led to three new optical identifications for the detected pulsars.  We 
find long-term optical periods for two of the pulsars, CXOU J010206.6-714115 
and SXP101, of 267 and 21.9 d, respectively.  Spectral analysis of a sub-set 
of the sample shows that the pulsars have harder spectra than the other 
sources detected.  By employing a quantile-based colour-colour analysis we 
are able to separate the detected pulsars from the rest of the sample.  Using 
archival catalogues we have been able to identify counterparts for the
majority of the sources in our sample.  Combining this with our results from 
the temporal analysis of the \chan\ data and archival optical data, the X-ray 
spectral analysis, and by determining the X-ray to optical flux ratios we
present preliminary classifications for the sources.  In addition to the four 
detected pulsars, our sample includes two candidate foreground stars, 12 
probable active galactic nuclei, and five unclassified sources.
\end{abstract}


\section{Introduction}

The Small Magellanic Cloud (SMC) is turning out to be an exciting nest of 
X-ray binary pulsars.  Estimates of the star formation rate (SFR) for the SMC 
range between $0.044 M_{\sun}$ yr$^{-1}$ from H$\alpha$ measurements 
\citep{ken91} to $0.38 M_{\sun}$ yr$^{-1}$ from supernova birth rates 
\citep{fil98}.  Using the relation between the X-ray luminosity function 
of high-mass X-ray binaries (HMXBs) and the SFR of the host galaxy 
\citep{gri03} and the upper and lower star formation rate estimates, 
\citet{sg05} predict between 6 and 49 HMXBs with luminosities $\geq 10^{35}$ 
erg s$^{-1}$ in the SMC.  We now know of $\sim 50$ such systems in the SMC 
\citep{hab04,coe05}.

Several of these detections have come from \chan\ and {\it RXTE} work over
the last couple of years \citep{edg04,lay05}.  This large number suggests a 
dramatic phase of star birth in the past, probably associated with the most 
recent closest approach between the SMC and the Large Magellanic Cloud 
\citep[LMC;][]{gar96}.  Even more extreme, \citet{naz03} analysed a 
$\sim 100$ ks exposure of just one $20\arcmin \times 20\arcmin$ \chan\ field
and identified more than 20 probable Be/X-ray binary systems.  Multiplying 
these numbers up by the $\sim2\degr \times 2\degr$ size of the SMC, and 
allowing for $\sim 10$\% X-ray duty cycles, suggests the final number of 
Be/X-ray binaries could be well in excess of 1,000.  Thus the study of the 
SMC is not only providing a great homogeneous sample of HMXBs for study, but 
is also providing direct insights into the history of our neighbouring galaxy.

\begin{figure}
 \includegraphics[width=84mm]{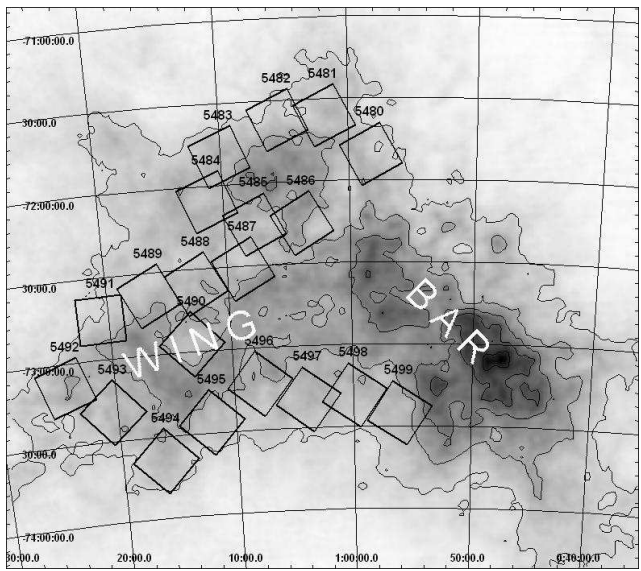}
 \caption{The location of the 20 fields studied by Chandra in this work, 
overlaid on a neutral hydrogen density image of the SMC \citep{sta99}.  The 
Wing and Bar of the SMC are marked.}
 \label{fig:smc_fields}
\end{figure}

\begin{table*}
 \centering
 \begin{minipage}{180mm}
  \caption{X-ray bright sources in the SMC Wing Survey.}
  \label{tab:src}
  \begin{tabular}{@{}rlcccccccc}
  \hline
   Object  &  Name  &  RA     &  Dec.   &  Error   &  Counts & 
   $P_{\rm pulse}$  & Pulsed fraction & Obs ID &  Date \\
           &        & (J2000) & (J2000) & (arcsec) &         &
    (s)             & (per cent)      &        &       \\      
 \hline
1 & CXOU J005551.5-733110 & 00:55:51.54 & -73:31:10.1 & 0.88 & 231 & -- & -- &
 5499 & 2006-03-03 \\
2 & RX J0057.3-7325 & 00:57:27.08 & -73:25:19.5 & 0.83 & 433 & 101.16 & 
 $29\pm7$ & 5499 & 2006-03-03 \\
3 & CXOU J005754.4-715630 & 00:57:54.41 & -71:56:30.9 & 0.95 & 130 & -- & -- & 
 5480 & 2006-02-06 \\
4 & CXOU J010014.2-730725 & 01:00:14.22 & -73:07:25.3 & 1.88 & 110 & -- & -- &
 5498 & 2006-03-03 \\
5 & CXOU J010206.6-714115 & 01:02:06.69 & -71:41:15.8 & 0.83 & 383 & 700.54 &  
 $35\pm9$ & 5481 & 2006-02-06 \\
6 & CXOU J010245.0-721521 & 01:02:45.01 & -72:15:21.7 & 0.91 & 81  & -- & -- & 
 5486 & 2006-02-10 \\
7 & SAX J0103.2-7209 & 01:03:13.94 & -72:09:14.4 & 0.90 & 244 & 337.51 & 
 $45\pm19$ & 5486 & 2006-02-10 \\
8 & CXOU J010455.4-732555 & 01:04:55.50 & -73:25:55.2 & 1.33 & 58  & -- & -- &
 5497 & 2006-03-03 \\
9 & CXOU J010509.6-721146 & 01:05:09.68 & -72:11:46.6 & 1.17 & 50  & -- & -- &
 5486 & 2006-02-10 \\
10 & CXOU J010533.0-721331 & 01:05:33.08 & -72:13:31.2 & 1.21 & 80  & -- & -- &
 5486 & 2006-02-10 \\
11 & CXOU J010712.6-723533 & 01:07:12.63 & -72:35:33.8 & 0.87 & 1919 & 65.78 & 
 $37\pm5$ & 5487 & 2006-02-10 \\
12 & CXOU J010735.0-732022 & 01:07:35.00 & -73:20:22.6 & 1.20 & 66  & -- & -- &
 5496 & 2006-03-03 \\
13 & CXOU J010836.6-722501 & 01:08:36.65 & -72:25:01.7 & 0.99 & 67  & -- & -- &
 5487 & 2006-02-10 \\
14 & CXOU J010849.5-721232 & 01:08:49.51 & -72:12:32.9 & 0.90 & 144 & -- & -- &
 5485 & 2006-02-08 \\
15 & CXOU J010855.6-721328 & 01:08:55.64 & -72:13:28.2 & 1.02 & 54  & -- & -- &
 5485 & 2006-02-08 \\
16 & CXOU J011021.3-715201 & 01:10:21.31 & -71:52:01.2 & 0.80 & 94  & -- & -- &
 5483 & 2006-02-06 \\
17 & CXOU J011050.6-721025 & 01:10:50.62 & -72:10:25.9 & 0.92 & 82  & -- & -- &
 5484 & 2006-02-06 \\
18 & CXOU J011154.2-723105 & 01:11:54.28 & -72:31:05.0 & 0.92 & 82  & -- & -- &
 5488 & 2006-02-12 \\
19 & CXOU J011303.4-724648 & 01:13:03.46 & -72:46:48.4 & 1.83 & 86  & -- & -- &
 5490 & 2006-02-27 \\
20 & CXOU J011744.7-733922 & 01:17:44.77 & -73:39:22.7 & 0.96 & 89  & -- & -- &
 5494 & 2006-03-01 \\
21 & CXOU J011832.4-731741 & 01:18:32.44 & -73:17:41.6 & 1.40 & 71  & -- & -- &
 5493 & 2006-02-27 \\
22 & CXOU J012027.3-724624 & 01:20:27.31 & -72:46:24.8 & 0.78 & 64  & -- & -- &
 5491 & 2005-07-24 \\
23 & CXOU J012223.6-730848 & 01:22:23.65 & -73:08:48.5 & 0.96 & 301 & -- & -- &
 5492 & 2005-08-12 \\
\hline								
\end{tabular}							
\end{minipage}							
\end{table*}							

It is important to note that, despite appearances, the SMC is a very
three dimensional object. Studies of the Cepheid population by \citet{lan86}
have revealed that the depth of the SMC is up to ten times its observed 
width.  The two main structures, the Bar and the Wing, lie $\sim11$ kpc 
behind, and $\sim8$ kpc in front of the main body of the SMC respectively. 

To date most of the X-ray studies by \chan\ and {\it RXTE} have concentrated on
the Bar which has proved to be a large source of HMXBs.  The \chan\ work 
presented here focuses on a study of the Wing of the SMC.  Its intention is to 
compare the pulsar population of the Wing with what we already know for the
Bar.  \citet{coe05} studied the locations of the detected X-ray pulsars
and believed they identified a relationship between the H{\sevensize I} 
intensity distribution and that of the pulsars.  They found that the pulsars
seem to lie in regions of low/medium H{\sevensize I} densities, suggesting
that high-mass star formation is well suited to these densities.  The 
locations of the \chan\ observations reported here are based upon this 
analysis.

\section{The {\it Chandra} SMC Wing Survey}

We performed a survey with \chan\ of the wings of the SMC between 2005 July 
and 2006 March (see Figure~\ref{fig:smc_fields}).  A total of 20 fields were 
observed using the standard ACIS-I imaging mode configuration which utilises
chips I0-I3 plus S2 and S3.  The exposure times for each field ranged 
from 8.6--10.3 ks.  We performed the data reduction using 
{\sevensize CIAO V}3.2.  The event files were filtered to restrict the energy 
range to 0.5--8.0 keV and we barycentrically corrected the data.  Potential 
sources were detected using the {\sevensize WAVDETECT} algorithm.  Here we 
present the analysis of the 23 brightest X-ray sources ($>50$ counts) in the 
survey (see Table~\ref{tab:src}).  The positional errors quoted are the 95\% 
confidence region for the source position that takes into account the 
properties of the telescope optics and the source brightness 
\citep[see][]{hong05}, combined in quadrature with the boresight error 
($\sim 0.7 \arcsec$ at 95\% confidence).  A full analysis and catalogue of 
sources from all 20 fields will be presented in a future publication 
(McGowan et al., in preparation).

\begin{figure*}
 \includegraphics[width=120mm]{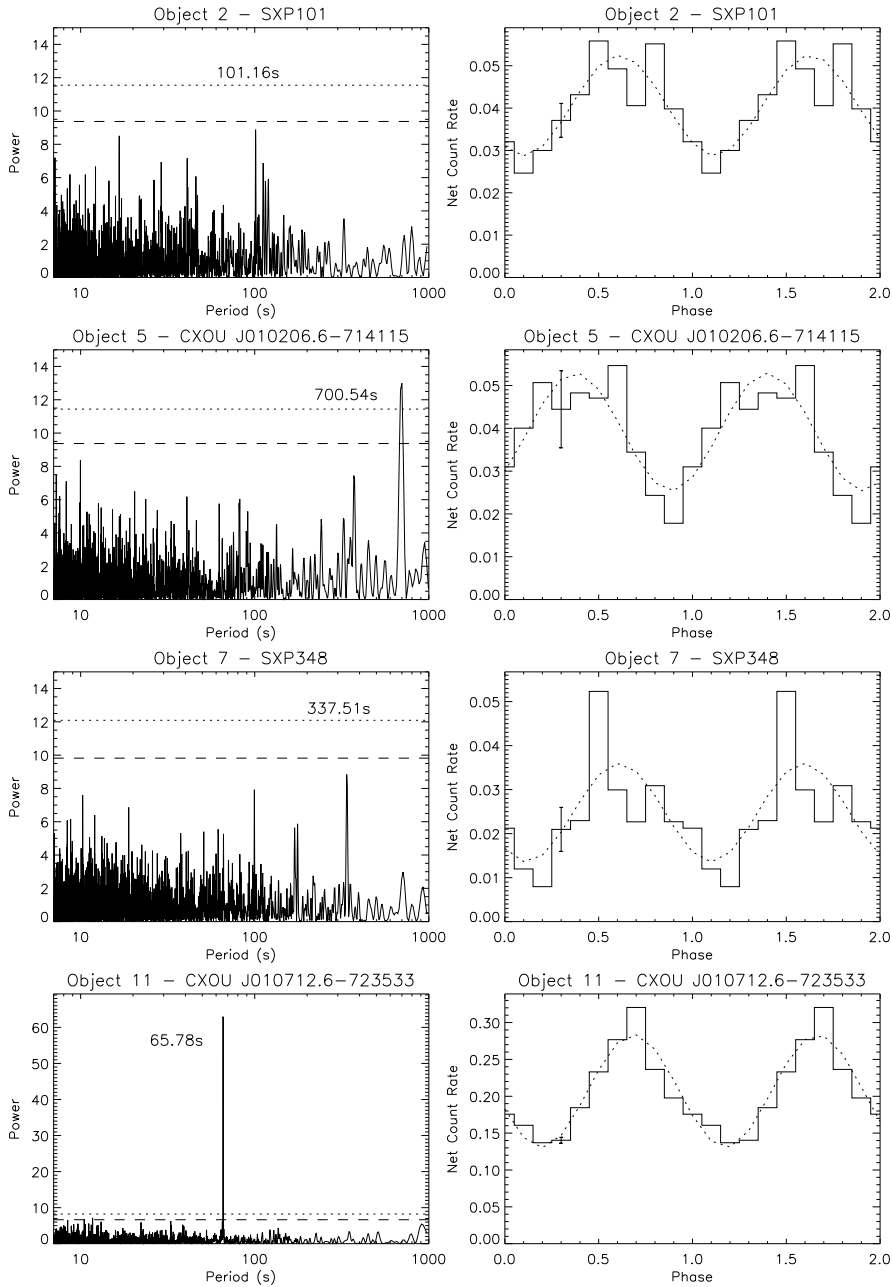}
 \caption{Temporal analysis of the \chan\ data.  Left: Lomb-Scargle 
periodograms for the sources where a significant peak was detected.  The 90\% 
(dashed line) and 99\% (dotted line) confidence limits are shown.  Right: the 
pulse profiles for the sources with error bars, where the uncertainty is the 
standard error for the data points in the bin.  The fitted sine functions are 
shown (dotted line).}
 \label{fig:pow_spec_fold}
\end{figure*}

\begin{table*}
 \centering
 \begin{minipage}{100mm}
  \caption{Spectral fits to the X-ray bright sources in the SMC Wing Survey.}
  \label{tab:spec}
  \begin{tabular}{@{}rlcccc}
  \hline
   Object & Model & $\Gamma$ & $T$        & $\Delta C$ (dof) & 
   $f_{X}$ \\
          &       &          & $10^{6}$ K &   & ergs cm$^{-2}$ s$^{-1}$ \\
 \hline
1 & PL & $1.9^{+0.2}_{-0.2}$ & ... & 0.5 (20) & $3.3 \times 10^{-13}$ \\
  & Brem & ... & $41.0^{+20.5}_{-12.0}$ & 0.4 (20) & $2.7 \times 10^{-13}$  \\
  & Mekal & ... & $49.4^{+20.0}_{-11.9}$ & 0.6 (20) & $2.9 \times 10^{-13}$ \\
2 & PL & $0.6^{+0.1}_{-0.1}$ & ... & 1.0 (35) & $1.0 \times 10^{-12}$ \\ 
  & Brem & ... & 2314 & 2.3 (35) & $7.5 \times 10^{-13}$ \\
  & Mekal & ... & 927 & 2.5 (35) & $7.2 \times 10^{-13}$ \\
3 & PL & $1.3^{+0.2}_{-0.2}$ & ... & 1.1 (55) & $2.5 \times 10^{-13}$ \\
  & Brem & ... & $273\pm 192$ & 1.1 (55) & $2.3 \times 10^{-13}$ \\
  & Mekal & ... & $270\pm 188$ & 1.1 (55) & $2.4 \times 10^{-13}$ \\ 
4 & PL & $3.4^{+0.4}_{-0.4}$ & ... & 0.8 (45) & $3.6 \times 10^{-13}$ \\ 
  & Brem & ... & $7.3^{+2.8}_{-1.4}$ & 0.8 (45) & $2.1 \times 10^{-13}$ \\
  & Mekal & ... & $14.9^{+1.6}_{-3.4}$ & 1.4 (45) & $1.2 \times 10^{-13}$ \\  
5 & PL & $0.4^{+0.1}_{-0.1}$ & ... & 0.8 (34) & $1.4 \times 10^{-12}$ \\
  & Brem & ... & 2314 & 3.2 (34) & $8.5 \times 10^{-13}$ \\ 
  & Mekal & ... & 927 & 3.6 (34) & $8.2 \times 10^{-13}$ \\ 
7 & PL & $0.8^{+0.2}_{-0.2}$ & ... & 1.4 (21) & $5.8 \times 10^{-13}$ \\ 
  & Brem & ... & 2314 & 2.0 (21) & $4.5 \times 10^{-13}$ \\
  & Mekal & ... & 927 & 2.2 (21) & $4.6 \times 10^{-13}$ \\
11 & PL & $0.3^{+0.1}_{-0.1}$ & ... & 1.1 (159) & $7.1 \times 10^{-12}$ \\
   & PL$^{a}$ & $0.5^{+0.1}_{-0.1}$ & ... & 1.1 (158) & $6.9 \times 10^{-12}$\\
   & Brem & ... & 2314 & 4.4 (159) & $4.0 \times 10^{-12}$ \\
   & Mekal & ... & 927 & 4.8 (159) & $4.4 \times 10^{-12}$ \\ 
14 & PL & $1.7^{+0.2}_{-0.2}$ & ... & 1.1 (59) & $2.4 \times 10^{-13}$ \\
   & Brem & ... & 133.4 & 2.0 (59) & $2.1 \times 10^{-13}$ \\ 
   & Mekal & ... & $74.1^{+88.5}_{-27.2}$ & 1.1 (59) & $2.3 \times 10^{-13}$ \\
23 & PL & $1.5^{+0.2}_{-0.2}$ & ... & 0.8 (105) & $5.2 \times 10^{-13}$ \\  
   & Brem & ... & $109.3^{+129.9}_{-41.8}$ & 0.8 (105) & $4.6 \times 10^{-13}$ \\  
   & Mekal & ... & $105.1^{+172.2}_{-39.0}$ & 0.9 (105) & $4.9 \times 10^{-13}$ \\
\hline								
\end{tabular}	

\medskip
The data were fitted with absorbed power-law (PL), bremsstrahlung (Brem) and
Mekal models, assuming a neutral hydrogen column density of $N_{\rm H} = 
0.06 \times 10^{22}$ cm$^{-2}$, where $\Gamma$ is the photon index and $T$ is
the temperature determined from the fits.  The goodness of fit using the 
C-statistic ($\Delta C$) and the unabsorbed flux in the 0.3--10 keV range are 
also given.  $^{a}$ Model fit results in $N_{\rm H} = 0.19^{+0.09}_{-0.08} 
\times 10^{22}$ cm$^{-2}$ \\

\end{minipage}							
\end{table*}							

\subsection{Temporal Analysis}
\label{sect:pulse}

The main goal of the SMC Wing Survey is to detect new pulsars.  We created 
background subtracted light curves for the 23 sources in our sample and 
searched for periodic variations from 6.5--1000 s.  The temporal analysis 
was performed using the Starlink {\sevensize PERIOD} software and we
generated Lomb-Scargle and Phase Dispersion Minimisation periodograms for 
each of our sources.

We determined the 90\% and 99\% confidence levels for the Lomb-Scargle 
periodograms from a cumulative probability distribution appropriate for each 
data set.  Using a Monte Carlo method we generated 10,000 simulated light 
curves with the same time sampling and variance as the real data.  The 
simulated light curves were taken from a Gaussian distribution.  A 
Lomb-Scargle periodogram was produced for each simulated light curve, and the 
peak power was recorded.  From these values the probability of obtaining a 
given peak power from pure noise can be calculated and the cumulative 
distribution function derived.

Two sources show a significant peak in the Lomb-Scargle periodogram and a 
corresponding strong peak in the Phase Dispersion Minimisation periodogram.
Two of the other sources in the sample are the previously known pulsars, 
RX J0057.3-7325 = SXP101 and SAX J0103.2-7209 = SXP348.  These two show 
pulsations, but at a level below the 90\% confidence limit.  The power 
spectra for all four sources are shown in Figure~\ref{fig:pow_spec_fold} 
(left).  We folded the light curves on the detected periods (see 
Figure~\ref{fig:pow_spec_fold}, right) and fitted the resulting pulse 
profiles with a sine function to determine the pulsed fractions (see 
Table~\ref{tab:src}).  We define the pulse fraction as $(F_{\rm max} - 
F_{\rm min}) / (F_{\rm max} + F_{\rm min})$, where $F_{\rm max}$ and 
$F_{\rm min}$ are the maximum and minimum of the fitted pulse light curve.

The error on the periods were determined using Kov\'acs formula 
\citep{kov81,hor86}.  Using this method the uncertainty calculated takes 
into account both the resolution due to the light curve sampling and the 
signal-to-noise ratio of the detected modulation.

\begin{table*}
 \centering
 \begin{minipage}{170mm}
  \caption{Optical counterparts for the X-ray bright sources in the SMC Wing 
Survey.}
  \label{tab:opt}
  \begin{tabular}{@{}rlllcl@{}}
  \hline
          & \multicolumn{3}{c}{Optical Counterpart} &  &  \\
   Object & OGLE III & OGLE II & MACHO & Period & Comment \\
          &          &         &       & (d)    &         \\    
 \hline
1 & SMC106.8 21521 & &  & - & Changes of $\sim 0.1$ mag over $\sim 400$ d \\
2 & SMC106.7 15343 & & 211.16415.11 & 21.9 & SXP101. Optical period seen in 
both OGLE III \\
   & & &             &      &  and MACHO data \\
3 & SMC108.4 384 &  &  & - & Some variability ($\sim 0.1$ mag) on timescales \\
  &              &  &  &   & of $\sim 100$ d \\
4 & & & 211.16591.6 & 29.6 & MACHO data only, saturated in OGLE III.  Period\\
   & & &             &      & is not strong in MACHO and affected by many \\
   & & &             &      & saturated points \\
5 & SMC114.7 39  &  &  & 267 & New pulsar, CXOU J010206.6-714115. Overall \\
  &              &  &  &   & brightness change of 0.5 mag over $\sim 1000$ d \\
6 &               & SMC-SC9 168928 & 206.16775.520 & - & No variability \\
7 & & SMC-SC9 173121 & 206.16776.17 & - & SXP348. Variability ($\sim0.1$ mag)\\
  &               &  &  &   & on timescales of $\sim 400$ d \\
8 & SMC111.7 8943 & &  & 1.49 & Sinusoidal folded light curve of 0.04 mag
amplitude \\ 
9 & SMC113.2 13190 & & 206.16947.35 &  - & No variability \\
10 & SMC113.2 13509 & & 206.16946.1089 & - & No variability \\
11 & &  SMC-SC11 48835 & 206.17055.21 &  - & New pulsar, CXOU J010712.6-723533.
Brightening \\
  &              &  &  &   & of $\sim 0.02$ mag over $\sim 1200$ d \\
12 & ? & & & ? & Not in OGLE or MACHO fields \\
13 & SMC113.1 9719 & SMC-SC11 116013 & 206.17114.1658 &  - & Brightening of 
$\sim 0.5$ mag over $\sim 800$ d \\
14 & SMC118.7 galaxy & SMC-SC11 120966 & 206.17175.739 & - & No variability \\
15 & SMC118.7 galaxy &  & 206.17174.524 & - & No variability \\
16 & SMC118.5 1160 &  &  & - & Some variability ($\sim 0.2$ mag) on 
timescales\\
 &               &  &  &   & of $\sim 100$ d \\
17 & SMC118.7 10314 &  &  & - & No variability \\
18 & SMC115.5 12842 & &  &  - & No variability \\
19 & SMC115.7 18128 & &  &  - & Brightening of $\sim 0.4$ mag in $\sim 1500$ 
d \\
20 & SMC117.4 3274 & &  &  - & Brightening of $\sim 0.4$ mag in $\sim 1000$ 
d \\
21 & SMC121.6 galaxy & &  & ? & Galaxy \\
22 & SMC120.7 6336 & &  &  - & No variability \\
23 & SMC121.4 542 & &  &  - & No variability \\
\hline								
\end{tabular}							
\end{minipage}							
\end{table*}							

\subsection{Spectral Analysis}

We have extracted spectra for the sources with $>100$ counts using 
{\sevensize CIAO V}3.2 tools.  The spectra were regrouped by requiring at 
least 10 counts per spectral bin for the brighter sources ($>200$ counts), 
and 2 counts per bin for the fainter sources ($100<$ counts $<200$).  The 
subsequent spectral fitting and analysis were performed using 
{\sevensize XSPEC V}12.3.0.  We fitted each spectrum with a power-law, 
thermal bremsstrahlung and Mekal model.  In each fit we included absorption.
Due to the low number of counts detected we fixed the column density at the 
value for the SMC of $6 \times 10^{20}$ cm$^{-2}$, only in the case of 
source \#11 (CXOU J010712.6-723533) have we also allowed the $N_{\rm H}$ to 
be fit.  The small number of detected counts renders chi-squared statistics 
invalid.  We have therefore used an alternative statistic, the C-statistic 
\citep{cas79}, for our model fitting.  The goodness of fit given in 
Table~\ref{tab:spec}, $\Delta C$, is determined in a similar way to reduced 
$\chi^{2}$.  The unabsorbed flux in the 0.3--10 keV band has been determined 
for each fit.  For sources that are members of the SMC, or assumed members 
(see Section \ref{sect:ratio}), we have also calculated the luminosity in the 
0.3--10 keV range, using a distance to the SMC of 60 kpc (based on the 
distance modulus \citet{wes97}).   The results of the spectral fitting 
are summarised in Table~\ref{tab:spec}.

\subsection{Long-term Optical Light Curves}

All 23 objects were investigated for possible optical periods (see 
Table~\ref{tab:opt}).  Such periods have been found in many previous HMXB 
systems in the SMC and can either represent the binary period of the system 
or non-radial pulsations (NRPs) from the mass-donor star 
\citep[for examples of both see][]{edg05,sch06}.  Data were collected from as 
many of the following possible sources: OGLE II \citep{szy05,uda97}, OGLE III 
and MACHO \citep{alc99}.  The OGLE III archive contains $I$-band photometry
spanning 5 years, the data are not yet public.  In each case the optical 
light curves were searched for periods in the ranges $10-1000$ d and 
$1-10$ d by generating Lomb-Scargle periodograms.  If a significant period 
was found, then the data were folded modulo that period with phase zero set 
to the time of maximum light phase.  In one case (source \#5, CXOU 
J010206.6-714115) the data were first detrended with a polynomial fit.  Quite 
a few of the objects revealed long-term variations on timescales of 
$100-1000$ d - such changes are similar to the common Type 4 variations 
reported by \citet{men02} from Be stars in the SMC.  The errors on the 
periods were determined using the same method as for the pulse periods (see 
Section \ref{sect:pulse}).

\subsection{X-ray to Optical Flux Ratios}
\label{sect:ratio}

Using the results from the spectral fitting and information from optical 
catalogues we have constructed X-ray to optical flux ratios for the majority 
of the sources in our sample (Table \ref{tab:ratio}).  It has been shown 
that such ratios are a good discriminator for different classes of sources 
(see e.g. \citealt{mac88,hor01}), with typical values for active galactic 
nuclei (AGN) lying in the region $\rm log (f_{X}/f_{opt}) = 0.0 \pm 1$.  We 
have used the flux in the 0.5--2 keV band and Eq. (3) from \citet{hor01} to 
calculate the ratio for each source that has a measured $R$ magnitude.  In 
the case of a source that has too few counts to model in XSPEC, we have 
determined the flux given the count rate and assuming a power-law index of 
1.6 (see Section \ref{sect:disc}) and neutral hydrogen column density of 
$6 \times 10^{20}$ cm$^{-2}$, using PIMMS v3.9a.  The results of this 
analysis allows us to provisionally classify the sources and determine 
whether the object is a member of the SMC (see Section \ref{sect:source}).

\subsection{Quantile Analysis}

We have used the quantile analysis technique of \citet{hong04} to investigate 
the X-ray colours of the 23 sources in our sample.  In a traditional hardness 
ratio the photons are split into predefined energy bands.  The quantile 
method divides the photon distribution into a given number of equal 
proportions, where the quantiles are the energy values that mark the 
boundaries between consecutive subsets.  This has the advantage, compared to 
traditional hardness ratios, that there is no spectral dependence and a 
colour can be calculated even for sources with very few counts 
\citep[for more details see][]{hong04}.

For each source we determine the median and quartiles of the photon 
energy distribution.  In Figure~\ref{fig:quantile} we show the quantile-based 
colour-colour diagram (QCCD), using the median and the ratio of two 
quartiles, for our sample.  In the diagram the spectrum hardens as one goes 
further right and changes from concave-downward to concave-upward moving from 
top to bottom \citep[see Figure 7,][]{hong04}.  We have included in the 
figure the pulsars detected by \citet{edg04}.  The new and known pulsars are 
marked.

\section{Individual Sources}
\label{sect:source}

\subsection{CXOU J005551.5-733110}  

This source was detected in observation ID 5499 taken on 2006 March 3 (MJD
53797).  The temporal analysis does not reveal any significant periodicities.
The location of CXOU J005551.5-733110 is close to source 65 in the {\it ROSAT} 
HRI catalogue of the SMC \citep{sas00}, being consistent within errors.
The position of the \chan\ source coincides with 2MASS 00555147-7331101 
($J=16.3$, $H=15.4$, $K=14.6$) and OGLE III source SMC106.8 21521 ($I=17.7$).
The optical light curve displays changes of $\sim 0.1$ mag over $\sim 400$ d.
The X-ray emission from the source is well fitted by a thermal bremsstrahlung 
with a temperature of $4.1 \times 10^{7}$ K, however the temperature is 
poorly constrained (see Table \ref{tab:spec}).  A power-law also fits the 
data well with an index of 1.9.  If this source is a member of the SMC the 
unabsorbed luminosities derived from these fits are $1.2 \times 10^{35}$ and 
$1.4 \times 10^{35}$ erg s$^{-1}$, respectively. 

\begin{table}
 \centering
 \begin{minipage}{80mm}
  \caption{X-ray to optical flux ratios.}
  \label{tab:ratio}
  \begin{tabular}{@{}rcccc}
  \hline
   Object & $R$ & $f_{X}$ & $\rm log (f_{X}/f_{R})$ & Class \\
 \hline
1 & - & $1.3 \times 10^{-13}$ & - & - \\ 
2 & 14.8 & $1.0 \times 10^{-13}$ & -1.58 & pulsar \\
3 & 20.1 & $8.3 \times 10^{-14}$ & 0.46 & AGN \\
4 & 10.9 & $1.3 \times 10^{-13}$ & -3.03 & star \\
5 & 15.1 & $9.7 \times 10^{-14}$ & -1.47 & pulsar \\
6 & 18.0 & $3.2 \times 10^{-14}$ & -0.79 & AGN \\ 
7 & 13.3 & $6.8 \times 10^{-14}$ & -2.34 & pulsar \\
8 & 14.5 & $2.8 \times 10^{-14}$ & -2.25 & star \\
9 & 18.0 & $2.0 \times 10^{-14}$ & -1.00 & - \\
10 & 18.7 & $3.2 \times 10^{-14}$ & -0.51 & AGN \\
11 & 14.9 & $4.4 \times 10^{-13}$ & -0.90 & pulsar \\
12 & - & $2.8 \times 10^{-14}$ & - & - \\
13 & 18.5 & $2.8 \times 10^{-14}$ & -0.65 & AGN \\
14 & 19.4 & $7.7 \times 10^{-14}$ & 0.15 & AGN \\
15 & 18.9 & $2.0 \times 10^{-14}$ & -0.64 & AGN \\
16 & 19.0 & $4.0 \times 10^{-14}$ & -0.30 & AGN \\
17 & - & $3.6 \times 10^{-14}$ & - & AGN \\
18 & 19.3 & $3.2 \times 10^{-14}$ & -0.27 & AGN \\
19 & - & $3.2 \times 10^{-14}$ & - & - \\
20 & 19.6 & $3.6 \times 10^{-14}$ & -0.10 & AGN \\
21 & - & $2.8 \times 10^{-14}$ & - & AGN \\
22 & - & $2.8 \times 10^{-14}$ & - & - \\
23 & 19.1 & $1.4 \times 10^{-13}$ & 0.29 & AGN \\
\hline	
\end{tabular}	
							
\medskip
$f_{X}$ is the unabsorbed flux in the 0.5--2 keV range.  The X-ray to optical 
flux ratio is determined using $\rm log (f_{X}/f_{R}) = \rm log f_{X} + 5.50 
+ R/2.5$  \citep[Eq. (3),][]{hor01}.\\

\end{minipage}							
\end{table}		

\subsection{RX J0057.3-7325 = SXP101}  

RX J0057.3-7325 = AX J0057.4-7325 \citep{kah99,tor00}, also known as SXP101 
\citep{coe05}, was detected in observation ID 5499 taken on 2006 March 3 
(MJD 53797).  Coherent pulsations with a period of $101.45 \pm 0.07$ s were 
detected in {\it ASCA} data \citep{tor00}.  The strongest peak in both period 
searches of the \chan\ data occurs at $101.16 \pm 0.86$ s (see Figure 
\ref{fig:pow_spec_fold}), however the peak in the Lomb-Scargle falls below 
the $<90\%$ confidence level and would not have been regarded as significant 
if the source was not already known.

\citet{edg03} narrowed the optical counterpart down to two sources, D and E in 
Table 2 in their paper.  Our more precise \chan\ position allows us to 
determine that the optical counterpart is source E, Table 2 in \citet{edg03}, 
the star MACS J0057-734 10 \citep{tuc96}.  This position is also consistent
with 2MASS 00572706-7325192 ($J=15.7$, $H=15.6$, $K=15.6$), MACHO object 
211.16415.11 ($V=14.9$, $R=14.8$) and OGLE III source SMC106.7 15343 
($I=15.6$).  We find a period of $21.94 \pm 0.10$ d in both the OGLE III 
and MACHO data (Figure~\ref{fig:opt_lc2}).  From the Corbet diagram 
\citep{cor86} we would expect a longer period for the source, perhaps twice 
the detected period, however there are no strong peaks in that region of the 
periodogram.  A period of 22.95 d is found in the X-ray data from {\it RXTE} 
with $T_{0} =$ MJD 2452111.4 (Galache et al., in preparation).  The spectrum 
of SXP101 is described well by a power-law with a photon index of 0.6.  The 
resulting unabsorbed luminosity is $4.3 \times 10^{35}$ erg s$^{-1}$.

\begin{figure*}
 \includegraphics[width=130mm]{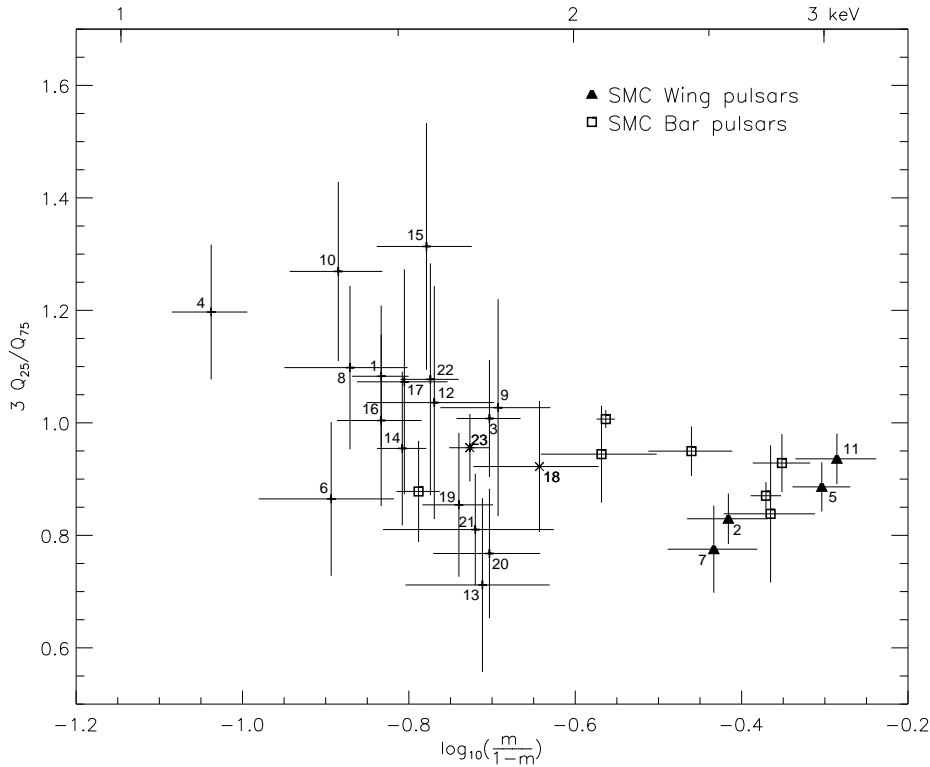}
 \caption{Quantile-based colour-colour diagram for the X-ray bright sources
in the SMC Wing survey.  The median and two quartiles of the photon energy 
distribution are given by $m$, $Q_{25}$ and $Q_{75}$, respectively.  The 
choice of x-axis allows the soft and hard phase space to be explored equally 
well \citep[for more details see][]{hong04}.  The new and known pulsars in
the SMC Bar \citep[open squares,][]{edg04} and the SMC Wing (filled triangles)
and marked.  Source \#6 is a known quasar, and source \#4 is a probable 
variable star.  The two sources marked with a cross are source \#18 
(CXOU J011154.2-723105) and source \#23 (CXOU J012223.6-730848).  Both of 
these sources display possible pulsations at 19.52 and 140.99 s, 
respectively, but their optical magnitudes and lack of long-term variability 
seems to rule out classification as Be/X-ray transients, and the nature of the 
sources are uncertain.}
 \label{fig:quantile}
\end{figure*}

\subsection{CXOU J005754.4-715630}  

This source was detected in observation ID 5480 taken on 2006 February 6 (MJD 
53772).  The temporal analysis does not reveal any significant periodicities.
Within errors the position of CXOU J005754.4-715630 is consistent with
USNO-B1.0 0180-0037995 ($B2=19.2$, $R2=20.1$), 2MASS J00575428-7156306
($J=16.8$, $H=16.4$, $K=14.8$) and OGLE III source SMC108.4 384 ($I = 18.5$).
The OGLE III light curve displays slight variation ($\sim 0.1$ mag) on 
timescales of $\sim 100$ d.  The X-ray spectrum can be well-fitted 
with a power-law with photon index of 1.3.  Statistically the data are equally
well-fitted with the thermal models, however the models are very poorly 
constrained (see Table \ref{tab:spec}).  The X-ray to optical flux ratio for 
this source implies that it is a background AGN.

\subsection{CXOU J010014.2-730725}  

This source was detected in observation ID 5498 taken on 2006 March 3 (MJD
53797).  The temporal analysis does not reveal any significant periodicities.

The position of CXOU J010014.2-730725 is close to the previously detected
source RX J0100.2-7307 \citep{kah99} and source 86 in the {\it ROSAT} HRI 
catalogue of the SMC \citep{sas00}.  The archival source has a 90\% confidence 
positional uncertainty of $10 \arcsec$ which encompasses the \chan\ position 
making it probable that they are the same source.  \citet{sas00} made a 
tentative classification of the source as a foreground star.  The counterpart 
to CXOU J010014.2-730725 is ISO-MCMS J010014.0-730725 detected by the 
Infrared Space Observatory and has been classified as a long-period red 
variable \citep{cio03}.  The source position is also consistent with 2MASS 
01001398-7307253 ($J=10.9$, $H=10.2$, $K=10.0$) and MACHO source 211.16591.6 
($V=12.2$, $R=10.9$).  The source is saturated in the OGLE III data.  
\citet{cio03} analysed the MACHO data for periodicities.  The blue light 
curve was considered of poor quality.  The red light curve was found to have 
a period of 29 d.  We have analysed the MACHO data, after removing all 
the saturated points we find a weak period at $29.57 \pm 0.35$ d in the red 
band (Figure~\ref{fig:opt_lc4}).  A larger peak at $367\pm 10$ d is too 
close to one year to be considered as astrophysical in origin.

The X-ray emission from CXOU J010014.2-730725 is very soft with almost no 
photons above $\sim 2$ keV.  The spectrum can be well-fitted with a 
power-law with index of 3.4 or a thermal bremsstrahlung with temperature 
$7.3 \times 10^{6}$ K (see Table \ref{tab:spec}).  The position of the source 
in the quantile-based colour-colour diagram suggests that it is a stellar 
coronal emission source \citep{hong05}.  Assuming this is the case, the X-ray 
to optical ratio based on the thermal fit is consistent with a galactic 
source.  The $V-R$ colour for the source is 1.3, consistent with an M star 
(M0-M2), as are the infrared colours.  We classify CXOU J010014.2-730725 as a 
foreground star.

\subsection{CXOU J010206.6-714115}  

Observation ID 5481 took place on 2006 February 6 (MJD 53772).  Timing 
analysis of this object revealed a period of $700.54 \pm 34.53$ s with a 
confidence of $>99$\% (see Figure \ref{fig:pow_spec_fold}).  The data were 
examined to ensure that this periodicity was not an artifact of the 707 s Z 
dithering frequency.

The position of this pulsar coincides with the emission-line star [MA93] 
1301 \citep{mey93}, the $V = 14.6$ mag O9 star AzV 294 \citep{mas02}, the 
OGLE III object SMC114.7 39 ($I = 14.3$), and 2MASS J01020668-7141161 
($J=14.2$, $H=14.0$, $K=13.9$).  The $B-V$ colour index is $-0.14$ 
\citep{mas02} which is consistent with the value expected from the optical 
companion of a Be X-ray binary.  The OGLE III light curve shows a 
strong period at $267.38 \pm 15.10$ d (Figure~\ref{fig:opt_lc5}).  This 
period is consistent with that predicted from the Corbet diagram 
\citep{cor86} for a 700 s Be/X-ray pulsar (following the convention of 
\citealt{coe05} this source would be designated SXP700).

The X-ray spectrum of CXOU J010206.6-714115 can be well-fitted with a 
power-law with index of 0.4 (see Table \ref{tab:spec}).  This model implies an 
unabsorbed luminosity of $6.0 \times 10^{35}$ erg s$^{-1}$.

\subsection{CXOU J010245.0-721521}  

This source was detected in observation ID 5486 taken on 2006 February 10 (MJD 
53776).  The position of CXOU J010245.0-721521 is consistent with OGLE II 
SMC-SC9 168928 ($B=19.4$, $V=18.9$ and $I=18.4$), a known quasar at $z=1.06$ 
\citep{dob03}.  The source is also detected by MACHO and is designated
206.16775.520 ($V=18.9$, $R=18.0$).  The optical light curves do not show any 
variability.  The position of the source on the QCCD indicates 
that the classification as a quasar is correct and the X-ray to optical flux 
ratio is consistent with a background AGN.

\begin{figure}
 \includegraphics[width=84mm]{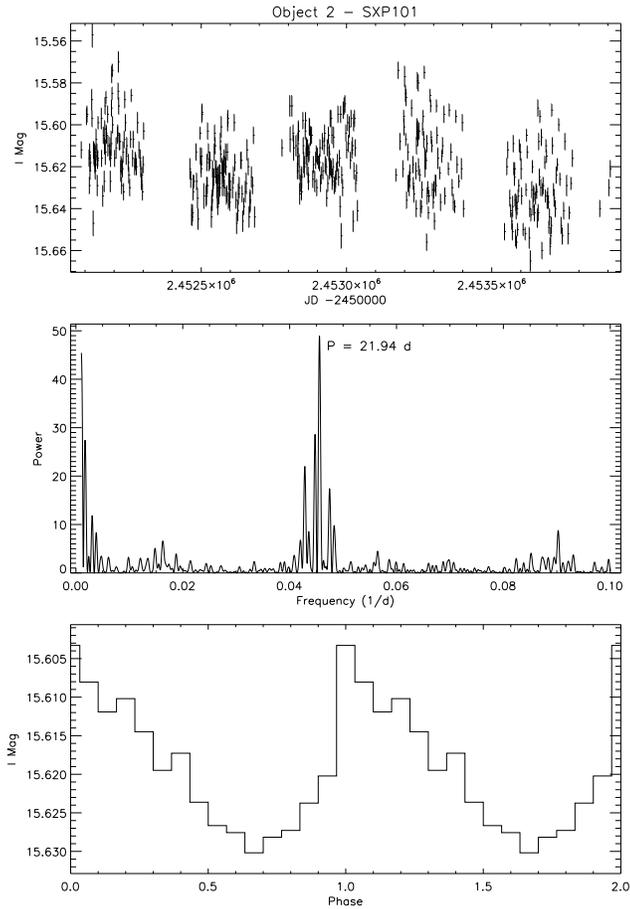}
 \caption{OGLE III light curve (top), Lomb-Scargle periodogram with 
significant period marked (middle) and folded light curve (bottom) for the 
optical counterpart of source \#2, RX J0057.3-7327 = SXP101.  The data have 
been folded on $P = 21.94$ d using $T_{0} =$ JD 2452124.8625.}
 \label{fig:opt_lc2}
\end{figure}

\begin{figure}
 \includegraphics[width=84mm]{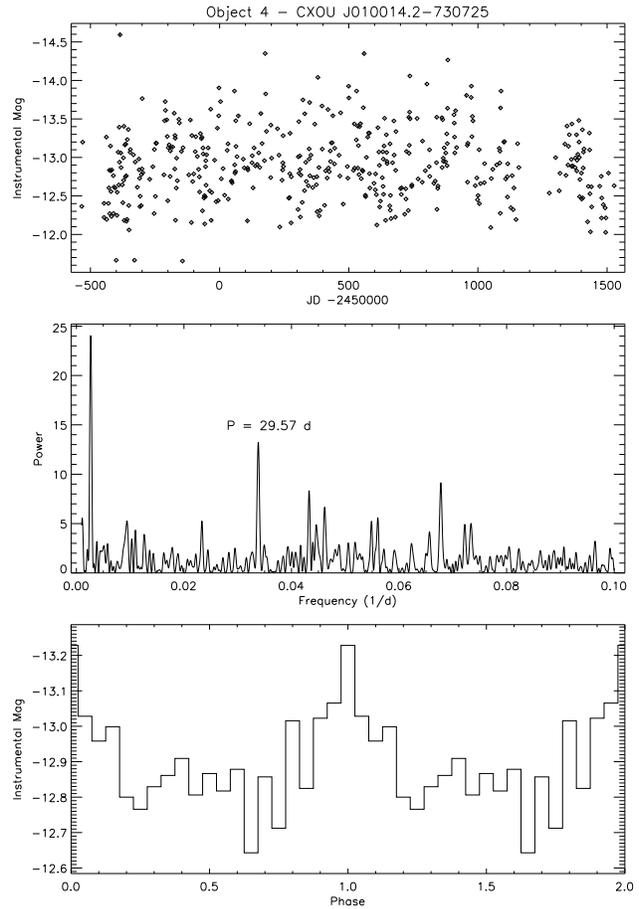}
 \caption{MACHO light curve (top), Lomb-Scargle periodogram with significant 
period marked (middle) and folded light curve (bottom) for the optical 
counterpart of source \#4, CXOU J010014.2-730725.  The data have been folded 
on $P = 29.57 $ d using $T_{0} =$ JD 2450322.5525.}
 \label{fig:opt_lc4}
\end{figure}

\subsection{SAX J0103.2-7209 = SXP348}  

SAX J0103.2-7209 = 2E 0101.5-7225 = RX J0103.2-7209 was first identified by 
{\it BeppoSAX} in 1998 \citep{isr98} and showed pulsations at a period of 
$345.2 \pm 0.1$ s.  The source, also known as SXP348 \citep{coe05}, was 
also seen in {\it ASCA} data taken in 1996, with a reported pulse period of 
$348.9 \pm 0.3$ s and period derivative (with respect to {\it BeppoSAX}) of 
1.7 s yr$^{-1}$ \citep{yok98}.  The pulse period determined from observations 
taken in 1999 with {\it Chandra} of $343.5 \pm 0.5$ s indicated that the 
pulsar had been spinning up with a constant rate since 1996 \citep{isr00}.  
Using serendipitous {\it XMM-Newton} observations of the source from 2000, 
\citet{hab04} measured a pulse period of $341.21 \pm 0.50$ s, implying that 
the spin-up was continuing.  

We observed SXP348 on 2006 February 10 (MJD 53776).  Although the highest 
peak in the Lomb-Scargle periodogram is at $337.51 \pm 5.17$ s, it falls below
the 90\% confidence limit (see Figure \ref{fig:pow_spec_fold}).  While our 
results seem to suggest that SXP348 may still be spinning-up, due to the 
uncertainty on the spin period we are unable to determine if the previously 
observed trend is still persisting.

The pulsar has previously been identified with a $V = 14.8$ mag Be star 
\citep{hug94,isr98}, [MA93] 1367 \citep{mey93}.  \citet{coe00} analysed OGLE 
II data of the proposed optical counterpart and concluded that it was the 
likely companion.  Timing analysis of the long-term $I$-band OGLE II data
did not show any periodic modulation in the range 1 to 50 d with an upper
limit of $\leq \pm 0.02$ mag.  The source position is also coincident with 
the MACHO object 206.16776.17 ($V=14.4$, $R=13.3$).  Our analysis of the 
light curve does not reveal any coherent period in the range 1 to 1000 d.
The source falls in a gap between the chips in OGLE III.

It has been shown that the spectrum of SXP348 can be well-fitted with an 
absorbed power-law with photon index $\sim 1.0$ \citep{isr98,yok98,hab04}.  
We find that the X-ray emission is well-fitted by a power-law with index 
0.8 (see Table \ref{tab:spec}).  This fit gives an unabsorbed luminosity of 
$2.5 \times 10^{35}$ erg s$^{-1}$.

\subsection{CXOU J010455.4-732555}  

This source was detected in observation ID 5497 taken on 2006 March 3 (MJD
53797).  The source is located very near to the edge of the chip and only
falls on the chip for part of the observation.  We were therefore unable to 
perform a search for pulsations.

The position of CXOU J010455.4-732555 is coincident with USNO-B1.0 
0165-0046936 ($R1=14.5$, $B2=15.3$, $R2=15.0$), 2MASS 01045550-7325558 
($J=12.4$, $H=11.8$, $K=11.8$) and OGLE III source SMC111.7 8943 ($I=13.0$).
The OGLE III light curve reveals a period of $1.4880 \pm 0.0005$ d 
(Figure~\ref{fig:opt_lc8}).  The next largest peak is at 3.0238 d.  It is
unclear whether the true period is 1.4880 or 3.0238 d, however the values are 
not consistent with the shorter period being a harmonic of the longer one.  The
peak at $\sim 3$ d could be due to the beating of the true period with the 1 
d sampling variation.  If the source was a Be/X-ray binary it is hard to 
reconcile the detected variation with an orbital period; in this case it 
could be a NRP.  However, the optical brightness of the object 
indicates that the source is likely to be a variable star.  The results from
the X-ray to optical flux ratio calculation are consistent with a Galactic 
foreground star.  Further X-ray observations to search for pulsations and 
follow-up optical spectroscopy are needed to determine the nature of this 
source.

\subsection{CXOU J010509.6-721146}  

This source was detected in observation ID 5486 taken on 2006 February 10 (MJD 
53776).  The temporal analysis does not reveal any significant periodicities.
The location of CXOU J010509.6-721146 lies within the $1\arcmin$ error 
circle of AX J0105-722 \citep{yok98}.  A study of the region around 
AX J0105-722 by \citet{fil00} resolved several sources.  \citet{fil00} 
proposed that the most likely counterpart to the {\it ASCA} source was RX 
J0105.1-7211.  Within errors the \chan\ source is also consistent with this 
{\it ROSAT} PSPC object.  A search for the optical counterpart to the 
{\it ASCA} source was carried out by \citet{coe05}.  Based on H$\alpha$ 
observations and temporal analysis of the H$\alpha$ and optical data, they 
identified the {\it ASCA} X-ray source, designated SXP3.34, with [MA93] 1506 
\citep{mey93}.  The position of this optical source also lies within the 
{\it ASCA} error circle but it is not consistent with the {\it ROSAT} or 
\chan\ source.  A search for pulsations could not be performed on the 
{\it ROSAT} data due to poor statistics, and we did not detect any 
significant pulsations in our search of the \chan\ data, therefore no firm
identification with AX J0105-722 can be made.  The \chan\ position coincides
with 2MASS 01050959-7211470 ($J=16.8$, $H=16.3$, $K=15.3$), MACHO object 
206.16947.35 ($V=18.2$, $R=18.0$) and OGLE III source SMC113.2 13190 
($I=17.7$).  The optical light curves do not show any changes.  The X-ray to 
optical flux ratio is -1.0, making it difficult to classify the source.

\subsection{CXOU J010533.0-721331}  

This source was detected in observation ID 5486 taken on 2006 February 10 (MJD 
53776).  The temporal analysis does not reveal any significant periodicities.

The position of CXOU J010533.0-721331 is close to RX J0105.5-7213
\citep{fil00}.  The statistical (90\%) positional error of $4\arcsec$ added 
in quadrature with the systematic error of $7\arcsec$ gives an overall 
positional error of $\sim 8\arcsec$ for RX J0105.5-7213.  This error circle 
encompasses CXOU J010533.0-721331, implying that they are the same source.
The position of the \chan\ source coincides with the radio counterpart, 
J0105.5-7213 \citep{fil00}, of the {\it ROSAT} source.  \citet{fil00} did not
detect any optical counterpart to this source and classified it as an 
optically faint background AGN.  However, we find 
that CXOU J010533.0-721331's position is consistent with MACHO object 
206.16946.1089 ($V=19.9$, $R=18.7$) and OGLE III object SMC113.2 13509 
($I=19.1$), and we propose that this source is the optical counterpart to 
the radio object.  There are no long-term variations in the optical light 
curves and the X-ray to optical flux ratio is consistent with a background
AGN.

\begin{figure}
 \includegraphics[width=84mm]{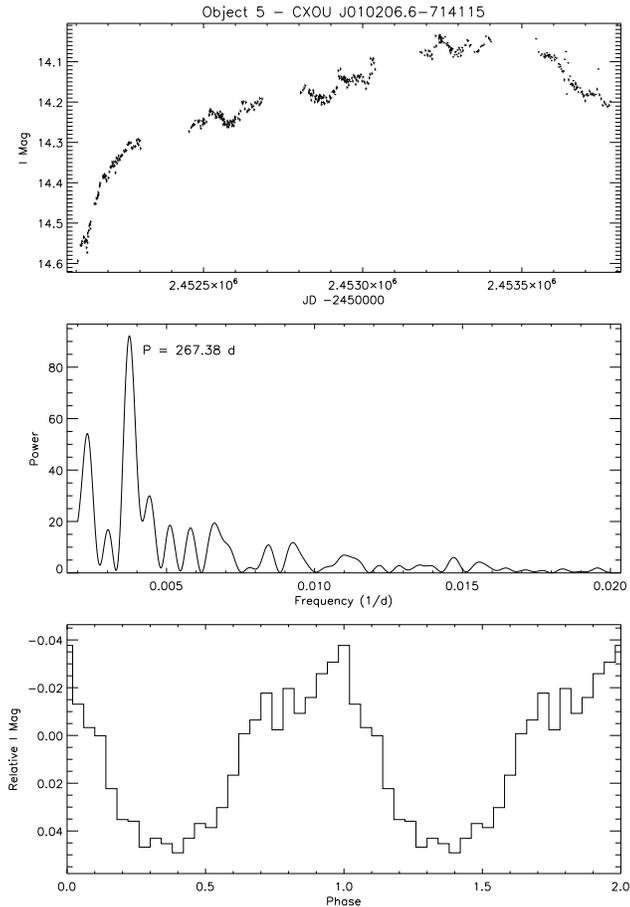}
 \caption{OGLE III light curve (top), Lomb-Scargle periodogram with 
significant period marked (middle) and folded light curve (bottom) for the 
optical counterpart of source \#5, CXOU J010206.6-714115.  The data 
were detrended using a polynomial before the period search was performed
and hence the folded light curve is given in relative $I$ mag.  The data 
have been folded on $P = 267.38 $ d using $T_{0} =$ JD 2452264.6099.}
 \label{fig:opt_lc5}
\end{figure}

\begin{figure}
 \includegraphics[width=84mm]{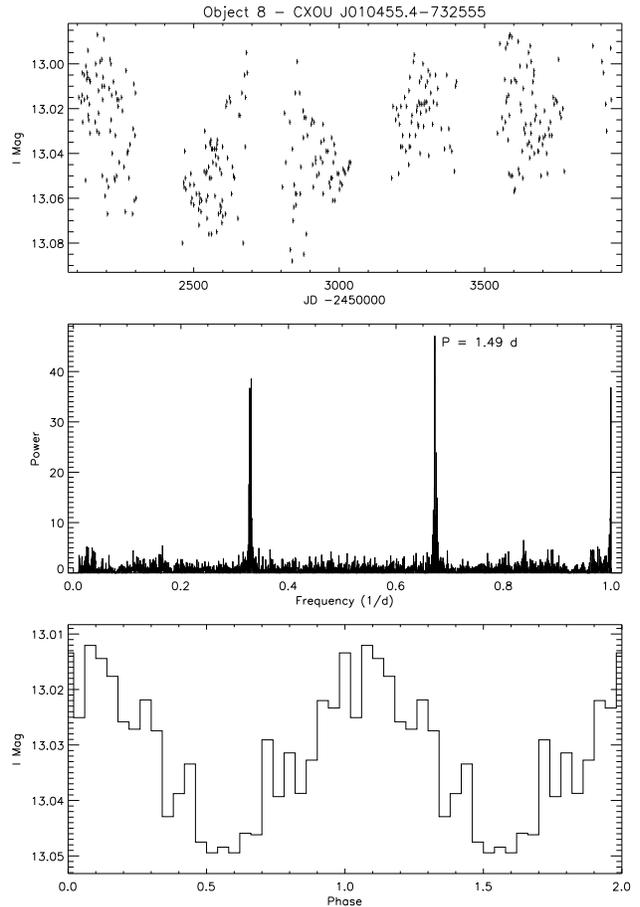}
 \caption{OGLE III light curve (top), Lomb-Scargle periodogram with 
significant period marked (middle) and folded light curve (bottom) for the 
optical counterpart of source \#8, CXOU J010455.4-732555.   The data have 
been folded on $P = 1.488 $ d using $T_{0} =$ JD 2452214.6844.}
 \label{fig:opt_lc8}
\end{figure}

\subsection{CXOU J010712.6-723533}  

This source was detected in observation ID 5487 taken on 2006 February 10 (MJD 
53776).  Timing analysis of this object revealed a period of $65.78 \pm 0.13$ 
s with a confidence of $>99$\% (see Figure \ref{fig:pow_spec_fold}).

The position of the source coincides with the emission line star [MA93] 1619
\citep{mey93}.  The source is also close to 2E 0105.7-7251 = RX J0107.1-7235
\citep{kah99}, taking into account the 90\% confidence positional uncertainty 
of $15\arcsec$ for this source, the position is consistent with CXOU 
J010712.6-723533 (following the convention of \citealt{coe05} this source 
would be designated SXP65.8).  2E 0105.7-7251 has been identified with a 
$V=16.6$ Be star, the $B-V$ colour index is -1.2.  CXOU J010712.6-723533's 
position is consistent with  2MASS 01071259-7235338 ($J=15.8$, $H=15.4$, 
$K=15.2$), MACHO object 206.17055.21 ($V=15.0$, $R=14.9$) and the OGLE II 
source SMC-SC11 48835 ($I=15.7$).  The long-term light curves show a 
brightening of $\sim 0.02$ mag over $\sim 1200$ d, but no periodic modulation 
is detected.  The source falls in a gap between the chips in OGLE III.  

We fitted the spectrum of CXOU J010712.6-723533 with a power-law and a 
blackbody, both with photoelectric absorption.  Initially we fixed the column 
density at $6 \times 10^{20}$ cm$^{-2}$.  This resulted in a best-fitting 
power-law with index 0.3 and unabsorbed luminosity of $3.1 \times 10^{36}$ 
erg s$^{-1}$.  Since we detected a high number of counts from this source, we 
also fitted the column density.  In this case the spectrum was again best fit 
with a power-law with index 0.5, $N_{\rm H}=1.9 \times 10^{21}$ cm$^{-2}$ and 
an unabsorbed luminosity of $3.0 \times 10^{36}$ erg s$^{-1}$ (see Table 
\ref{tab:spec}).  Statistically the power-law fits cannot be distinguished 
from each other.

\subsection{CXOU J010735.0-732022}  

This source was detected in observation ID 5496 taken on 2006 March 3 (MJD
53797).  The temporal analysis does not reveal any significant periodicities.
There are no optical or infrared matches for this source.

\subsection{CXOU J010836.6-722501}  

This source was detected in observation ID 5487 taken on 2006 February 10 (MJD 
53776).  The temporal analysis does not reveal any significant periodicities.
The position of CXOU J010836.6-722501 is consistent within errors to the 
position of source 117 in the {\it ROSAT} HRI SMC catalogue \citep{sas00}.  
The spectrum of the source was found to be hard \citep{hab00} and led 
\citet{sas00} to classify the source as an X-ray binary or AGN.  The position 
of the \chan\ source also coincides with MACHO object 206.17114.1658 
($V=18.8$, $R=18.5$), OGLE II source SMC-SC11 116013 ($I=19.5$) and OGLE III 
source SMC113.1 9719 ($I=19.6$).  The long-term optical light curve shows a 
brightening of $\sim 0.5$ mag over $\sim 800$ d.  As we do not detect 
pulsations from the source, and taking into account its position on the 
QCCD compared to the pulsars we have detected and its X-ray to optical flux 
ratio, we conclude it is likely that this source is a background AGN.

\subsection{CXOU J010849.5-721232}  

This source was detected in observation ID 5485 taken on 2006 February 8 (MJD
53774).  The temporal analysis does not reveal any significant periodicities.
The position of CXOU J010849.5-721232 is consistent with the MACHO object
206.17175.739 ($V=20.3$, $R=19.4$), the OGLE II object SMC-SC11 120966 
($I=20.0$) and the OGLE III object SMC118.7 galaxy ($I=20.5$).  The long-term 
optical light curves do not vary.  The X-ray emission from CXOU 
J010849.5-721232 can be described by a power-law with a photon index of 1.7 
or with a thermal emission model with temperature $7.4 \times 10^{7}$ K (see 
Table \ref{tab:spec}).  The temperature is poorly constrained and based on 
this the non-thermal fit is preferred.  In this case the X-ray to optical 
ratio indicates that the source is a background AGN.

\subsection{CXOU J010855.6-721328}  

This source was detected in observation ID 5485 taken on 2006 February 8 (MJD
53774).  The temporal analysis does not reveal any significant periodicities.
The position of CXOU J010855.6-721328 coincides with OGLE III source 
SMC118.7 galaxy and MACHO object 206.17174.524 ($V=19.5$, $R=18.9$).  There
are no changes apparent in the optical light curves and the X-ray to optical 
flux ratio is consistent with a background AGN.

\subsection{CXOU J011021.3-715201}  

This source was detected in observation ID 5483 taken on 2006 February 6 (MJD 
53772).  The temporal analysis does not reveal any significant periodicities.
Within errors the position of CXOU J011021.3-715201 is consistent with 
USNO-B1.0 0181-0039286 ($R1=19.0$, $B2=18.8$, $R2=19.1$) and OGLE III source 
SMC118.5 1160 ($I=18.8$).  The long-term optical light curve shows some 
variation ($\sim 0.2$ mag) on $\sim 100$ d timescales.  The X-ray to optical 
flux ratio implies that the source is a background AGN.

\subsection{CXOU J011050.6-721025}  

This source was detected in observation ID 5484 taken on 2006 February 6 (MJD 
53772).  The temporal analysis does not reveal any significant periodicities.
Within errors the position of CXOU J011050.6-721025 is coincident with the 
radio source [FBR2002] J011050-721026 \citep{fil02} and OGLE III source 
SMC118.7 10314 ($I=20.5$).  Analysis of the long-term OGLE III light curve 
does not show any variations.  Based on the identification of the \chan\ source
with a radio object suggests CXOU J011050.6-721025 is a background AGN.

\subsection{CXOU J011154.2-723105}  

This source was detected in observation ID 5488 taken on 2006 February 12 (MJD
53778).  A peak at $19.52 \pm 0.03$ s with $>90\%$ confidence was detected in
the Lomb-Scargle periodogram.  A search using Phase Dispersion Minimisation
does not find a similar modulation.

A tentative detection of an X-ray source, designated BKGS 20, $\sim 3\arcsec$
from CXOU J011154.2-723105 has previously been reported \citep{bru87}.  
However, a more stringent analysis of the {\it Einstein} data by 
\citet{wan92} failed to detect the source.  The position of the \chan\ source 
is coincident with USNO-B1.0 0174-0065503 ($R1=19.3$, $B2=18.9$, $R2=18.5$) 
and the OGLE III source SMC115.5 12842 ($I=19.1$).  

The position of the source on the quantile analysis plot is intriguing
(Figure \ref{fig:quantile}); it seems to lie between the detected pulsars
and the rest of the sources in our sample.  The source shows possible 
pulsed emission, however the magnitudes of the optical counterpart, and its 
lack of variability on long timescales, indicates that it is not a Be/X-ray 
transient.  In addition, the X-ray to optical flux ratio indicates that the 
source is a background AGN.

\subsection{CXOU J011303.4-724648}  

This source was detected in observation ID 5490 taken on 2006 February 27 (MJD
53793).  The temporal analysis does not reveal any significant periodicities.
The position of CXOU J011154.2-723105 is consistent with OGLE III source
SMC115.7 18128 ($I=19.8$), whose light curve exhibits a brightening over 
$\sim 1500$ d of $\sim 0.4$ mag.

\subsection{CXOU J011744.7-733922}  

This source was detected in observation ID 5494 taken on 2006 March 1 (MJD 
53795).  The temporal analysis does not reveal any significant periodicities.
A {\it ROSAT} PSPC X-ray source has previously been detected near the 
position of CXOU J011744.7-733922.  The archival source, [HFP2000] 537 
\citep{hab00}, has a positional uncertainty of $\sim 10\arcsec$ which 
encompasses the position of the \chan\ source.  This suggests that they are 
the same object.  CXOU J011744.7-733922's position is consistent with 
USNO-B1.0 0163-0044338 ($R1=19.6$, $B2=19.5$, $R2=19.2$) and OGLE III source 
SMC117.4 3274 ($I=18.4$).  The OGLE III light curve shows a brightening of 
$\sim 0.4$ mag in $\sim 1000$ d.  The X-ray to optical flux ratio implies
that the source is a background AGN.

\subsection{CXOU J011832.4-731741}  

This source was detected in observation ID 5493 taken on 2006 February 27 (MJD
53793).  The temporal analysis does not reveal any significant periodicities.
A {\it ROSAT} PSPC X-ray source has previously been detected near to the 
position of CXOU J011832.4-731741.  The archival source, [HFP2000] 449 
\citep{hab00}, has a positional uncertainty of $\sim 9\arcsec$ which 
encompasses the position of the \chan\ source.  It is therefore likely that 
they are the same object.  CXOU J011832.4-731741's position coincides with
OGLE III source SMC121.6 galaxy, which suggests that it is a background AGN.

\subsection{CXOU J012027.3-724624}  

This source was detected in observation ID 5491 taken on 2005 July 24 (MJD
53575).  The temporal analysis does not reveal any significant periodicities.
The position of CXOU J012027.3-724624 coincides with OGLE III source
SMC120.7 6336 ($I=20.5$).  The long-term optical light curve does not show
any variability.

\subsection{CXOU J012223.6-730848}  

This source was detected in observation ID 5492 taken on 2005 August 12 (MJD
53594).  A peak with $> 90\%$ confidence was detected in the Lomb-Scargle
periodogram at $140.99 \pm 1.50$ s, however there is no corresponding strong 
peak in the Phase Dispersion Minimisation periodogram.  Analysis of the data
with the method used for finding pulsars in {\it RXTE} data \citep{gal06} 
reveals a possible periodicity at 282.49 s, approximately twice the value 
from the Lomb-Scargle periodogram.  In this method the light curve is folded 
on the period found in the Lomb-Scargle to produce a pulse profile.  The 
pulse profile is then used as a template to subtract the pulsations from the 
light curve.  A Lomb-Scargle periodogram is generated for the cleaned light 
curve, and this power spectrum is subtracted from the original.  The resulting
power spectrum shows only the contribution of the pulsar to the Lomb-Scargle,
allowing possible harmonics that may have been lost in the noise to be 
detected.

The position of CXOU J012223.6-730848 is consistent with USNO-B1.0 
0168-0053065 ($R1=19.1$, $B2=20.8$, $R2=19.3$) and OGLE III source SMC121.4 
542 ($I=18.9$).  There are no changes detected in the optical light curves. 
The X-ray spectrum is described equally well by a power-law with photon index 
1.5 and a bremsstrahlung with a temperature of $1.1 \times 10^{8}$ K (see 
Table \ref{tab:spec}).  The temperature is poorly constrained and we therefore
prefer the non-thermal fit to the data.  

The optical magnitude, lack of long-term variability in the optical light 
curve and the position of the source on the QCCD (see Figure 
\ref{fig:quantile} and Section 4) suggests that this source may not be a 
pulsar.  The X-ray to optical flux ratio determined using the non-thermal 
model fit implies that the source is a background AGN.  Further observations 
are needed to verify the existence of pulsations and determine the nature of 
the source.

\section{Discussion}
\label{sect:disc}

Observations have shown that the SMC contains a large number of HMXBs, 
providing a homogeneous sample with which to investigate the evolution of the 
SMC, and to compare with our Galaxy.  The HMXBs are tracers of very young 
populations and the SMC seems to have a particularly prominent young 
population.  To date, studies of the SMC have mostly concentrated on the Bar 
of the SMC.  To obtain a full picture of the history of the SMC we need to 
broaden our study and include the outer regions of the SMC.

In this first paper from the SMC Wing Survey we have investigated the X-ray 
and optical characteristics of the 23 brightest X-ray sources.  The temporal 
analysis, combined with identification of the optical counterparts, shows 
that our sample contains four pulsars, two newly detected and two previously 
known.  The statistical significance of this cannot be determined until a 
full analysis of the entire SMC Wing Survey has been performed.  The other 
sources include a quasar and possibly two foreground stars.  The 
classification of the remainder are not conclusive, but the lack of 
pulsations, long-term periodic variability, optical identifications and X-ray 
to optical flux ratios imply that they are most likely background AGN.  The 
classifications of several of the sources from the literature would seem to 
agree with this.  If the preliminary classifications we have presented in 
this paper are confirmed our results indicate that the spectral hardness and 
quantile analysis could be used to distinguish between different classes of
object (see below).

We have analysed the spectra of the 11 sources that have $>100$ counts (see 
Table \ref{tab:spec}).  Apart from source \#4, which is probably a variable 
star, most of the objects exhibit non-thermal emission.  We have been able to 
fit the spectra of all four pulsars in our sample, their power-law indices 
display a limited range of values with an average of 0.5 and standard 
deviation of 0.2.  In comparison, the spectra of the remaining seven sources 
with $>100$ counts have softer spectra, with an average photon index of 
$1.6\pm 0.2$ (excluding \#4).  \citet{hab04} found that the distribution
of photon indices for SMC HMXBs, mainly located in the Bar, had an average
of $1.0\pm 0.2$.  This is in agreement with the Bar pulsars studied by 
\citet{edg04} that have an average photon index of $1.1\pm 0.5$.  In general, 
the pulsars that we have detected in the Wing have harder spectra than those 
in the Bar.

It is likely that the pulsars have a higher intrinsic neutral hydrogen column 
density than the AGN, however it is puzzling why the Wing pulsars as a
group are harder/more absorbed than the Bar group.  Could the Wing pulsars 
be situated at the back of the SMC?  The work of \citet{lan86} would seem to 
rule out that possibility as they found that the Wing lies in front of the 
main body of the SMC.  It is more likely that small number statistics are
contributing to the observed division of sources.

A large fraction of our sources are too faint to extract a meaningful 
spectrum.  By constructing a quantile-based colour-colour diagram we 
have been able to investigate the spectral properties of all 23 sources in 
our sample (see Figure~\ref{fig:quantile}).  Our analysis shows that the 
pulsars we have detected in the Wing fall in a distinct group on the QCCD.  
The location of the Bar pulsars from \citet{edg04} also seem to fall in the 
harder part of the diagram, but the separation of sources is less clearly 
defined, with one source falling amongst the softer sources from the Wing.  
There does not seem to be anything remarkable about this particular Wing 
pulsar.  The softer sources include an identified star, quasar and possible 
AGN.  The source that appears to sit in the transition region between the 
majority of the pulsars and the other sources is CXOU J011154.2-723105 
(source \#18).  This object displays a possible pulsation of 19.52 s, but its 
optical magnitude and lack of long-term variability seems to rule out a 
Be/X-ray transient, and the nature of the source remains unclear.  

The classification of all of the sources will require optical spectroscopy, 
but the QCCD may be a useful tool for distinguishing pulsars from other types 
of object (stars, quasars, AGN) for the fainter X-ray sources in the SMC Wing 
survey.

\section{Summary}

We have detected two new pulsars, CXOU J010712.6-723533 and CXOU 
J010206.6-714115, and observed two previously known pulsars, SXP101 and 
SXP348.  With the accurate positions provided by \chan\ we have been able to 
determine new optical identifications for the two new pulsars, CXOU 
J010712.6-723533 and CXOU J010206.6-714115, and SXP101.  We have found 
long-term optical periods of 267 d and 21.9 d for CXOU J010206.6-714115 and 
SXP101, respectively.

\section*{Acknowledgments}

AU was supported by the BST grant of the Polish MNSW.  RHDC and SL acknowledge
support from Chandra/NASA grant GO5-6042A/NAS8-03060.
This research has made use of the SIMBAD data base, operated by CDS, 
Strasbourg, France.  This paper utilizes public domain data originally 
obtained by the MACHO Project, whose work was performed under the joint 
auspices of the U.S. Department of Energy, National Nuclear Security 
Administration by the University of California, Lawrence Livermore National 
Laboratory under contract No. W-7405-Eng-48, the National Science Foundation 
through the Center for Particle Astrophysics of the University of California 
under cooperative agreement AST-8809616, and the Mount Stromlo and Siding 
Spring Observatory, part of the Australian National University.

\label{lastpage}


\begin{thebibliography}{99}

\bibitem[\protect\citeauthoryear{Alcock et al.}{1999}]{alc99}
Alcock C., et al., 1999, PASP, 111, 1539
\bibitem[\protect\citeauthoryear{Corbet}{1986}]{cor86} Corbet R.H.D., 1986,
MNRAS, 220, 1047
\bibitem[\protect\citeauthoryear{Bruhweiler et al.}{1987}]{bru87}
Bruhweiler F.C., Klinglesmith D.A. III, Gull T.R., Sofia S., 1987, ApJ, 317, 
152
\bibitem[\protect\citeauthoryear{Cash}{1979}]{cas79}
Cash, W. 1979, ApJ, 228, 939
\bibitem[\protect\citeauthoryear{Cioni et al.}{2003}]{cio03}
Cioni M.-R.L., et al., 2003, A\&A, 406, 51
\bibitem[\protect\citeauthoryear{Coe \& Orosz}{2000}]{coe00}
Coe M.J., Orosz J.A., 2000, MNRAS, 311, 169
\bibitem[\protect\citeauthoryear{Coe et al.}{2005}]{coe05}
Coe M.J., Edge W.R.T., Galache J.L., McBride V.A., 2005, MNRAS, 356, 502
\bibitem[\protect\citeauthoryear{Dobrzycki et al.}{2003}]{dob03}
Dobrzycki A., Macri L.M., Stanek K.Z., Groot P.J., 2003, AJ, 125, 1330
\bibitem[\protect\citeauthoryear{Edge \& Coe}{2003}]{edg03}
Edge W.R.T., Coe M.J., 2003, MNRAS, 338, 428
\bibitem[\protect\citeauthoryear{Edge et al.}{2004}]{edg04}
Edge W.R.T., Coe M.J., Galache J.L., McBride V.A., Corbet R.H.D., 
Markwardt C.B., Laycock S., 2004, MNRAS, 353, 1286
\bibitem[\protect\citeauthoryear{Edge et al.}{2005}]{edg05}
Edge W.R.T., Coe M.J., Galache J.L., McBride V.A., Corbet R.H.D., 
Okazaki A.T., Laycock S., Markwardt C.B., Marshall F.E., Udalski A., 2005,
MNRAS, 361, 743
\bibitem[\protect\citeauthoryear{Filipovic et al.}{1998}]{fil98}
Filipovic M.D., 1998, A\&AS, 127, 119
\bibitem[\protect\citeauthoryear{Filipovic et al.}{2000}]{fil00}
Filipovic M.D., Haberl F., Pietsch W., Morgan D.H., 2000, A\&A, 353, 129
\bibitem[\protect\citeauthoryear{Filipovic et al.}{2002}]{fil02}
Filipovic M.D., Bohlsen T., Reid W., Staveley-Smith L., Jones P.A., Nohejl K.,
Goldstein G., 2002, MNRAS, 335, 1085
\bibitem[\protect\citeauthoryear{Galache}{2006}]{gal06}
Galache J.L., 2006, PhD Thesis, 78
\bibitem[\protect\citeauthoryear{Gardiner \& Noguchi}{1996}]{gar96}
Gardiner L.T., Noguchi M., 1996, MNRAS, 278, 191
\bibitem[\protect\citeauthoryear{Grimm, Gilfanov \& Sunyaev}{2003}]{gri03}
Grimm, H.-J., Gilfanov M.R., Sunyaev R.A., 2003, MNRAS, 339, 793
\bibitem[\protect\citeauthoryear{Haberl et al.}{2000}]{hab00}
Haberl F., Filipovic M.D., Pietsch W., Kahabka P., 2000, A\&AS, 142, 41
\bibitem[\protect\citeauthoryear{Haberl \& Pietsch}{2004}]{hab04} 
Haberl F., Pietsch W., 2004, A\&A, 414, 667
\bibitem[\protect\citeauthoryear{Hong et al.}{2004}]{hong04} 
Hong J., Schlegel E.M., Grindlay J.E., 2004, ApJ, 614, 508
\bibitem[\protect\citeauthoryear{Hong et al.}{2005}]{hong05} 
Hong J., van den Berg M., Schlegel E.M., Grindlay J.E., Koenig  X., 
Laycock S., Zhao P., 2005, ApJ, 635, 907
\bibitem[\protect\citeauthoryear{Horne \& Baliunas}{1986}]{hor86}
Horne J. H., Baliunas S. L., 1986, ApJ, 302, 757
\bibitem[\protect\citeauthoryear{Hornschemeier et al.}{2001}]{hor01}
Hornschemeier A.E., et al., 2001, ApJ, 554, 742
\bibitem[\protect\citeauthoryear{Hughes \& Smith}{1994}]{hug94}
Hughes J.P., Smith R.C., 1994, AJ, 107, 4
\bibitem[\protect\citeauthoryear{Israel et al.}{1998}]{isr98}
Israel G.L., Stella L., Campana S., Covino S., Ricci D., Oosterbroek T., 1998,
IAUC 6999, 1
\bibitem[\protect\citeauthoryear{Israel et al.}{2000}]{isr00}
Israel G.L., et al., 2000, ApJ, 531, 131
\bibitem[\protect\citeauthoryear{Kahabka et al.}{1999}]{kah99}
Kahabka P., Pietsch W., Filipovic M.D., Haberl F., 1999, A\&AS, 136, 81
\bibitem[\protect\citeauthoryear{Kennicutt}{1991}]{ken91}
Kennicutt R.C., Jr., 1991, in Haynes R.F., Milne D.K., eds., Proc. IAU Symp. 
148, The Magellanic Clouds. Reidel, Dordrecht, p. 139
\bibitem[\protect\citeauthoryear{Kov\'acs}{1981}]{kov81}
Kov\'acs G., 1981, Ap\&SS, 78, 175
\bibitem[\protect\citeauthoryear{Laney \& Stobie}{1986}]{lan86} 
Laney C.D., Stobie R.S., 1986, MNRAS, 222, 449
\bibitem[\protect\citeauthoryear{Laycock et al.}{2005}]{lay05}
Laycock S., Corbet R.H.D., Coe M.J., Marshall F.E., Markwardt C., 
Lochner J., 2005, ApJS, 161, 96
\bibitem[\protect\citeauthoryear{Maccacaro et al.}{1988}]{mac88}
Maccacaro T., Gioia I.M., Wolter A., Zamorani G., Stocke J.T., 1988, ApJ, 326, 
680
\bibitem[\protect\citeauthoryear{Massey}{2002}]{mas02} 
Massey P., 2002, ApJS, 141, 81
\bibitem[\protect\citeauthoryear{Mennickent et al.}{2002}]{men02}
Mennickent R.E., Pietrzynski G., Gieren W., Szewczyk O., 2002, A\&A, 393, 887
\bibitem[\protect\citeauthoryear{Meyssonnier \& Azzopardi}{1993}]{mey93}
Meyssonnier N., Azzopardi M., 1993, A\&AS, 102, 451
\bibitem[\protect\citeauthoryear{Naz\'e et al.}{2003}]{naz03}
Naz\'e Y., Hartwell J.M., Stevens I.R., Manfroid J., Marchenko S., 
Corcoran M.F., Moffat A.F.J., Skalkowski G., 2003, ApJ, 586, 983
\bibitem[\protect\citeauthoryear{Sasaki et al.}{2000}]{sas00}
Sasaki M., Haberl F., Pietsch W., 2000, A\&AS, 147, 75
\bibitem[\protect\citeauthoryear{Schmidtke \& Cowley}{2006}]{sch06}
Schmidtke P.C., Cowley A.P., 2006, AJ, 132, 919
\bibitem[\protect\citeauthoryear{Shtykovskiy \& Gilfanov}{2005}]{sg05}
Shtykovskiy P., Gilfanov M., 2005, MNRAS, 362, 879
\bibitem[\protect\citeauthoryear{Stanimirovi\'c et al.}{1999}]{sta99}
Stanimirovi\'c S., Staveley-Smith L., Dickey J.M., Sault R.J., Snowden S.L., 
1999, MNRAS, 302, 417
\bibitem[\protect\citeauthoryear{Szyma\' nski}{2005}]{szy05}
Szyma\' nski M.K., 2005, AcA, 55, 43
\bibitem[\protect\citeauthoryear{Torii et al.}{2000}]{tor00}
Torii K., Kohmura T., Yokogawa J., Koyama K., 2000, IAUC, 7441, 2 
\bibitem[\protect\citeauthoryear{Tucholke et al.}{1996}]{tuc96}
Tucholke H.-J., De Boer K.S., Seitter W.C., 1996, A\&AS, 119, 91
\bibitem[\protect\citeauthoryear{Udalski et al.}{1997}]{uda97}
Udalski A., Kubiak M., Szyma\' nski M., 1997, AcA, 47, 319
\bibitem[\protect\citeauthoryear{Wang \& Wu}{1992}]{wan92}
Wang Q., Wu X., 1992, ApJS, 78, 391
\bibitem[\protect\citeauthoryear{Westerlund}{1997}]{wes97}
Westerlund B., 1997, The Magellanic Clouds, Cambridge Univ. Press, Cambridge
\bibitem[\protect\citeauthoryear{Yokogawa \& Koyama}{1998}]{yok98}
Yokogawa J., Koyama K., 1998, IAUC 7009, 3

\end{thebibliography}
\end{document}